\documentclass[runningheads]{article}

\usepackage{graphicx} 

\usepackage{amsmath,amssymb}                      

\def\R{\ifmmode{I\hskip -3pt R}
           \else{\hbox{$I\hskip -3pt R$}}\fi}
\def\argmin{\mathop{\rm arg\,min}\limits}
\def\argmax{\mathop{\rm arg\,max}\limits}

\def\CV{{\rm CV}}

\title{Cluster validation by measurement of clustering characteristics 
relevant to the user}

\author{
Christian Hennig\\
Department of Statistical Science,\\ 
University College London,\\
Gower Street,\\ London WC1E 6BT, United Kingdom\\
(Email: {\tt c.hennig@ucl.ac.uk})
}

\begin{document}
\maketitle             
\setlength{\leftskip}{0pt}
\setlength{\headsep}{16pt}
\begin{abstract}
There are many cluster analysis methods that can produce quite 
different clusterings on the same dataset. Cluster validation is about the 
evaluation of the quality of a clustering; ``relative cluster validation'' 
is about using such criteria to compare clusterings. This can be used to 
select one of a set of clusterings from different methods, or from the same
method ran with different parameters such as different numbers of clusters. 

There are many cluster validation indexes in the literature. Most of them
attempt to measure the overall quality of a clustering by a single number, but 
this can be inappropriate. There are various different characteristics of a 
clustering that can be relevant in practice, depending on the aim of clustering,
such as low within-cluster distances and high between-cluster separation. 

In this paper, a number of validation criteria will be introduced that refer to
different desirable characteristics of a clustering, and that characterise
a clustering in a multidimensional way. In specific applications the user may 
be interested in some of these criteria rather than others. A focus of the 
paper is on methodology to standardise the different characteristics so that
users can aggregate them in a suitable way specifying weights for the various
criteria that are relevant in the clustering application at hand.
~
\end{abstract}

\section{Introduction}
The aim of the present paper is to present a range of cluster validation indexes that provide a multivariate assessment covering different complementary aspects of cluster validity. Here I focus on ``internal'' validation criteria that measure the quality of a clustering without reference to external information such as a known ``true'' clustering. Furthermore I am mostly interested in comparing different clusterings on the same data, which is often referred to as ``relative'' cluster validation. This can be used to select one of a set of clusterings from different methods, or from the same method ran with different parameters such as different numbers of clusters. 

In the literature (for an overview see Halkidi \textit{et al.}\cite{HVH16}) many cluster validation indexes are proposed. Usually these are advertised as measures of global cluster validation in a univariate way, often under the implicit or explicit assumption that for any given dataset there is only a single best clustering. Mostly these indexes are based on contrasting a measure of within-cluster homogeneity and a measure of between-clusters heterogeneity such as the famous index proposed by Calinski and Harabasz\cite{CH74}, which is a standardised ratio of the traces of the pooled within-cluster covariance matrix and the covariance matrix of between-cluster means.   

\begin{figure}
\centerline{
\includegraphics[width=125pt]{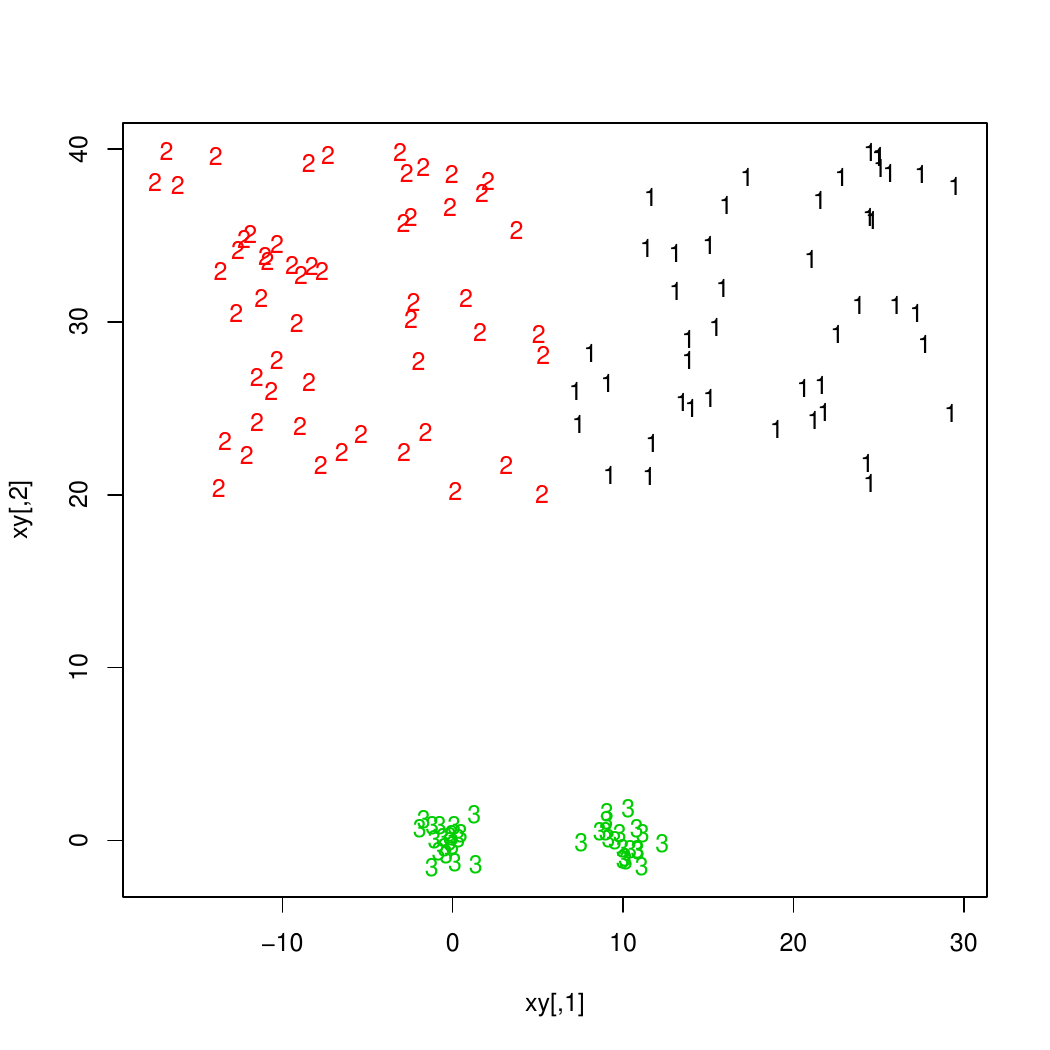}
\includegraphics[width=125pt]{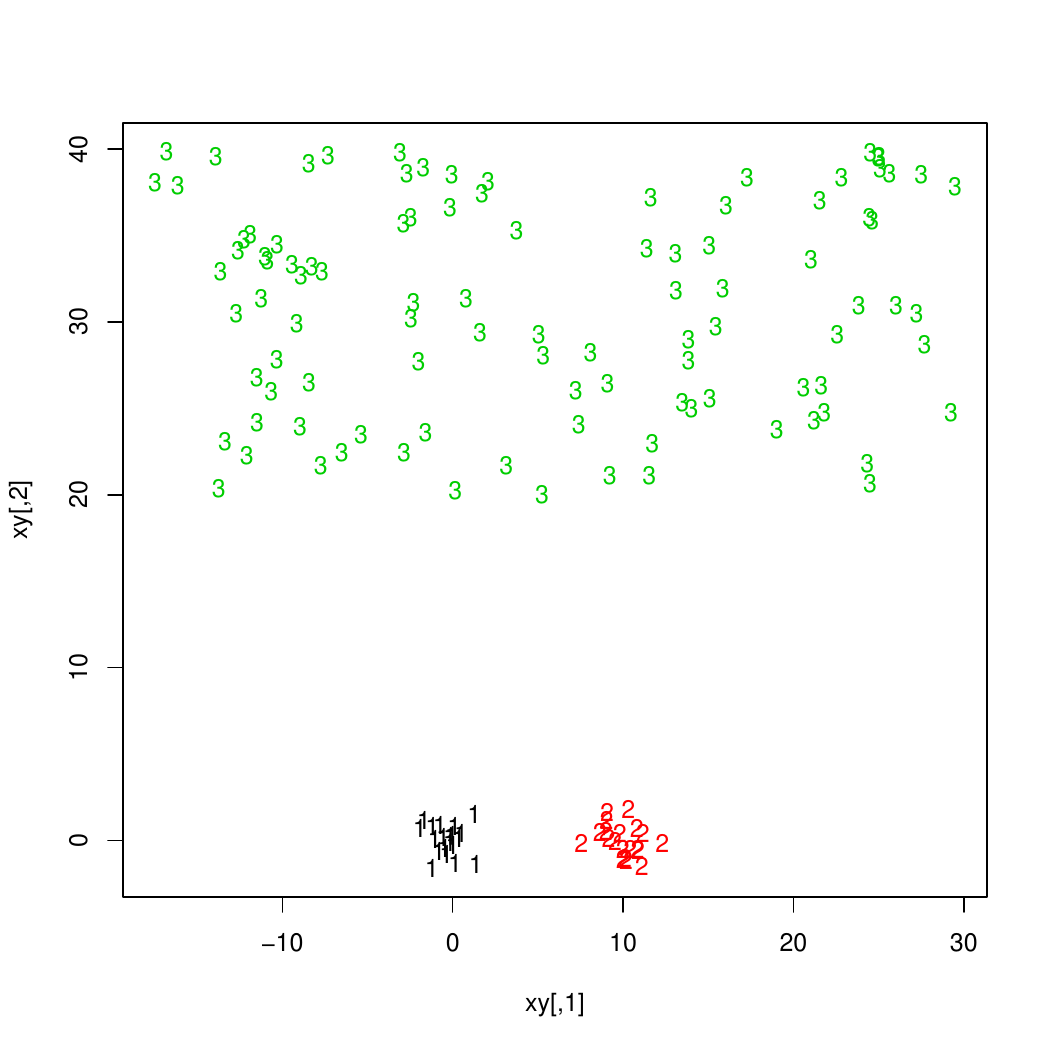}}
\caption{Artificial dataset. Left side: Clustering by 3-means.
Right side: clustering by Single Linkage with 3 clusters.}
\label{fxyunif}
\end{figure}

In Hennig\cite{Hen16} (see also Hennig\cite{Hen15}) I have argued that depending on the subject-matter background and the clustering aim different clusterings can be optimal on the same dataset. For example, clustering can be used for data compression and information reduction, in which case it is important that all data are optimally represented by the cluster centroids; or clustering can be used for recognition of meaningful patterns, which are often characterised by clear separating gaps between them. In the former situation, large within-cluster distances are not desirable, whereas in the latter situation large within-cluster distances may not be problematic as long as data objects occur with high density and without gap between the objects between which the distance is large. See Figure \ref{fxyunif} for two different clusterings on an artificial dataset with 3 clusters that may be preferable for these two different clustering aims.

Given a multivariate characterisation of the validity of a clustering, for a given application a user can select weights for the different characteristics depending on the clustering aim and the relevance of the different criteria. A weighted average can then be used to choose a clustering that is suitable for the specific application. This requires that the criteria measuring different aspects of cluster validity and normalised in such a way that their values are comparable when doing the aggregation. Although it is easy in most cases to define criteria in such a way that their value range is $[0,1]$, this is not necessarily enough to make their values comparable, because within this range the criteria may have very different variation. The idea here is that the expected variation of the criteria can be explored using resampled random clusterings (``stupid K-centroids'', ``stupid nearest neighbour clustering'') on the same dataset, and this can be used for normalisation and comparison.

The approach presented here can also be used for benchmarking cluster analysis 
methods. Particularly, it does not only allow to show that methods are better or
worse on certain datasets, it also allows to characterise the specific strength
and weaknesses of clustering algorithms in terms of the properties of the
found clusters.

Section \ref{snotation} introduces the general setup and defines notation.
In Section \ref{saspects}, all the indexes measuring different relevant aspects 
of a clustering are presented. Section \ref{sagg} defines an aggregated index
that can be adapted to practical needs. The indexes cannot be 
suitably aggregated in their raw form, and Section \ref{sstupid} introduces
a calibration scheme using randmly generated clusterings. Section 
\ref{sexamples} applies the methodology to two datasets, one illustrative 
artificial one and a real dataset regarding species delimitation. Section
\ref{sconc} concludes the paper. 

\section{General notation}\label{snotation}
Generally, cluster analysis is about finding groups in a set of objects
${\cal D}=\{x_1,\ldots,x_n\}$. There is much literature
in which the objects $x_1,\ldots,x_n$ are assumed to be
from Euclidean space $\R^p$, but in principle the could be from any
space ${\cal X}$. 

A clustering is 
a set ${\cal C}=\{C_1,\ldots,C_K\}$ with 
$C_j\subseteq {\cal D},\ j=1,\ldots,K$. The number of clusters $K$ may be
fixed in advance or not. For $j=1,\ldots,K$, let $n_j=|C_j|$ be the number
of objects in $C_j$. 
Obviously not every such ${\cal C}$ qualifies as a ``good'' or ``useful'' 
clustering, but what is demanded of ${\cal C}$ differs in
the different approaches of cluster analysis. Here ${\cal C}$ is required
to be a partition, e.g., $j\neq k \Rightarrow C_j \cap C_k=\emptyset$ and 
$\bigcup_{j=1}^K C_j={\cal D}$. For partitions, let 
$\gamma:\ \{1,\ldots,n\}\mapsto \{1,\ldots,K\}$ be the assignment function, 
i.e., $\gamma(i)=j$ if $x_i\in C_j$.
Some of the indexes introduced below could
also by applied to clusterings that are not partitions (particularly objects 
that are not a member of any cluster could just be ignored), but this is not
treated here to keep things simple. Clusters are here also assumed to be crisp
rather than fuzzy,
i.e., an object is either a full member of a cluster or not a member of this
cluster at all. In case of probabilistic clusterings, which give as output
probabilities $p_{ij}$ for object $i$ to be member of cluster $j$, it is assumed
that objects are assigned to the cluster $j$ maximising $p_{ij}$; in case
of hierarchical clusterings it is assumed that the hierarchy is cut at a 
certain number of clusters $K$ to obtain a partition.

Most of the methods introduced here are based on dissimilarity data.  A 
dissimilarity is a function  
$d:\ {\cal X}^2\mapsto \R^+_0$ so that
$d(x,y)=d(y,x)\ge 0$ and 
$d(x,x)=0$ for $x,y\in{\cal X}$.
Many dissimilarities are distances, i.e., they also fulfil the triangle
inequality, but this is not necessarily 
required here. Dissimilarities are extremely flexible, they can be defined
for all kinds of data, such as functions, time series, categorical data, 
image data, text data etc. If data are Euclidean, obviously
the Euclidean distance can be used. See Hennig\cite{Hen16} for a more general
overview of dissimilarity measures used in cluster analysis. 

\section{Aspects of cluster validity} \label{saspects}
In this Section I introduce measurements for various aspects of cluster 
validity. 
\subsection{Small within-cluster dissimilarities}
A major aim in most cluster analysis applications is to find homogeneous 
clusters. This often means that all
the objects in a cluster should be very similar
to each other, although it can in principle also have different meanings,
e.g., that a homogeneous probability model (such as the Gaussian distribution,
potentially with large variance) can account for all observations in a cluster.

The most straightforward way to formalise that all objects within a
cluster should be similar to each other is the average within-cluster distance,
although this needs to be weighted for cluster sizes so that every observation
has the same contribution to it:
\begin{displaymath}
I_{withindis}({\cal C}) = \frac{1}{n}\sum_{j=1}^K\frac{2}{n_j-1}
\sum_{x\neq y\in C_j}d(x,y).
\end{displaymath}
Smaller values are better. Knowing the data but not the clustering, the minimum possible value of $I_{withindis}$ is zero and the maximum is
$d_{max}=\max_{x,y\in{\cal D}}d(x,y)$, so 
$I_{withindis}^*({\cal C})=1-\frac{I_{withindis}({\cal C})}{d_{max}}\in[0,1]$ is a normalised version. When different criteria are aggregated (see Section \ref{sagg}), it is useful to define them in such a way that they point in the same direction; I will define all normalised indexes so that larger values are better. For this reason $\frac{I_{withindis}({\cal C})}{d_{max}}$ is subtracted from 1.

There are alternative ways of measuring whether within-cluster dissimilarities are overall small. All of these operationalise cluster homogeneity in slightly different ways. The 
objective function of K-means clustering can be written down as a constant times the average of all squared within-cluster Euclidean distances (or more general dissimilarities), which is an alternative measure, giving more emphasis to the biggest within-cluster dissimilarities. Most radically, one could use the maximum within-cluster dissimilarity. On the other hand one could use quantiles or trimmed means in order to make the index less sensitive to large within-cluster dissimilarities, although I believe that in most applications in which within-cluster similarity is important, these should be avoided and the index should therefore be sensitive against them. 
\subsection{Between-cluster separation}
Apart from within-cluster homogeneity, the separation between clusters is most often taken into account in the literature on cluster validation (most univariate indexes balance separation against homogeneity in various ways). Separation as it is usually understood cannot be measured by averaging all between-cluster dissimilarities, because it refers to what goes on ``between'' the clusters, i.e., the smallest between-cluster dissimilarities, whereas the dissimilarities between pairs of farthest objects from different clusters should not contribute to this.

The most naive way to measure separation is to use the minimum between-cluster dissimilarity. This has the disadvantage that with more than two clusters it only looks at the two closest clusters, and also in many applications there may be an inclination to tolerate the odd very small distance between clusters if by and large the closest points of the clusters are well separated. 

I propose here an index that takes into account a portion $p$, say $p=0.1$, of objects in each cluster that are closest to another cluster.

For every object $x_i\in C_j,\ i=1,\ldots,n,\ j\in\{1,\ldots,K\}$ let
$d_{j:i}=\min_{y\not\in C_j}d(x_i,y)$. Let 
$d_{j:(i)}\le\ldots \le d_{j:(n_j)}$ be the values of $d_{j:i}$ for $x_i\in C_j$ ordered from the smallest to the largest, and let $\lfloor pn_j\rfloor$ be the largest integer $\le pn_j$. Then the $p$-separation index is defined as 
\begin{displaymath}
 I_{p-sep}({\cal C})=\frac{1}{\sum_{j=1}^K \lfloor pn_j\rfloor}\sum_{j=1}^K
\sum_{i=1}^{\lfloor pn_j\rfloor} d_{j:(i)}.
\end{displaymath}
Obviously, $I_{p-sep}({\cal C})\in [0,d_{max}]$ and large values are good, therefore $I_{p-sep}^*({\cal C})=\frac{I_{p-sep}({\cal C})}{d_{max}}\in[0,1]$ is a suitable normalisation. 

\subsection{Representation of objects by centroids}
In some applications clusters are used for information reduction, and one way of
doing this is to use the cluster centroids for further analysis rather than the
full dataset. It is then relevant to measure how well the observations in a 
cluster are represented by the cluster centroid. The most straightforward method
to measure this is to average the dissimilarities of all objects to the 
centroid of the cluster they're assigned to. Let 
$c_1,\ldots,c_K$ be the centroids of clusters 
$C_1,\ldots, C_K$. Then,
\begin{displaymath}
  I_{centroid}({\cal C})=\frac{1}{n}\sum_{i=1}^n d(x_i, c_{\gamma(i)}).
\end{displaymath}
Some clustering methods such as K-means and Partitioning 
Around Medoids (PAM, Kaufman and Rousseeuw\cite{KR90}) are centroid-based, i.e.,
they compute the cluster centroids along with the clusters. Centroids can also 
be defined for the output of non-centroid-based methods, most easily as
\begin{displaymath}
  c_j=\argmin_{x\in C_j}\sum_{\gamma(i)=j} d(x_i, 
x),
\end{displaymath}
which corresponds to the definition of PAM. Again, there are possible 
variations. K-means uses squared Euclidean distances, and in case of Euclidean
data the cluster centroids do not necessarily have to be members of 
${\cal D}$,
they could also be mean vectors of the observations in the clusters. 

Again, by definition, $I_{centroid}({\cal C})\in[0,d_{max}]$. Small values are
better, and therefore $I_{centroid}^*({\cal C})=1-\frac{I_{centroid}({\cal C})}{d_{max}}\in[0,1]$.

\subsection{Representation of dissimilarity structure by clustering}
Another way in which the clustering can be used for information reduction is that the clustering can be seen as a more simple summary or representation of the dissimilarity structure. This can be measured by correlating the vector of pairwise dissimilarities ${\bf d}={\rm vec}\left([d(x_i,x_j)]_{i<j}\right)$
with the vector of a ``clustering induced dissimilarity'' 
${\bf c}={\rm vec}\left([c_{ij}]_{i<j}\right)$, where $c_{ij}=1(\gamma(i)\neq\gamma(j))$, and $1(\bullet)$ denotes the indicator function. With $r$ denoting the sample 
Pearson correlation,
  \begin{displaymath}
    I_{Pearson\Gamma}({\cal C})=r({\bf d},{\bf c}).
  \end{displaymath}
This index goes back to Hubert and Schultz\cite{HS76}, see also
Halkidi \textit{et al.}\cite{HVH16} for alternative versions.  
$I_{Pearson\Gamma}\in[-1,1]$, and large values are good, so it can be normalised
by $I^*_{Pearson\Gamma}=\frac{I_{Pearson\Gamma}+1}{2}\in[0,1]$.

\subsection{Small within-cluster gaps}
The idea that a cluster should be homogeneous can mean that there are no 
``gaps'' within a cluster, and that the cluster is well connected. A gap 
can be characterised as a split of a cluster into two subclusters so that the
minimum dissimilarity between the two subclusters is large. The 
corresponding index measures the ``length'' (dissimilarity) of the widest 
within-cluster gap (an alternative would be to average widest gaps 
over clusters):
\begin{displaymath}
  I_{widestgap}({\cal C})=\max_{C\in {\cal C}, D, E:\ C=D\cup E}\min_{x\in D, y\in E}d(x,y).  
\end{displaymath}
$I_{widestgap}\in[0,d_{max}]$ and low values are good, so it is normalised as 
$I^*_{widestgap}=1-\frac{I_{widestgap}}{d_{max}}\in[0,1]$.

A version of this taking into account density values is defined in Section 
\ref{sdens}. Widest gaps can be found computationally 
by constructing the within-cluster minimum spanning trees; the widest
distance occurring there is the widest gap. 
\subsection{Density modes and valleys}\label{sdens}
A very popular idea of a cluster is that it corresponds to a density mode, and
that the density within a cluster goes down from the cluster mode to the outer
regions of the cluster. Correspondingly, there should be density valleys 
between different clusters.

The definition of indexes that measure such a behaviour is based on a density
function $h$ that assigns a density value $h(x)$ to every observation. For
Euclidean data, standard density estimators such as kernel density estimators 
can be used. For general dissimilarities, I here propose a simple kernel density
estimator. Let $q_{d,p}$ be the $p$-quantile of the vector of dissimilarities 
${\bf d}$, e.g., for $p=0.1$, the 10\% smallest dissimilarities are 
$\le q_{d,0,1}$. Define the kernel and density as 
\begin{displaymath}
\kappa(d)=\left(1-\frac{1}{q_{d,p}}d\right)1(d\le q_{d,p}),\qquad
h(x)=\sum_{i=1}^n \kappa(d(x,x_i)).
\end{displaymath}
These can be normalised to take a maximum of 1:
\begin{displaymath}
h^*(x)=\frac{h(x)}{\max_{y\in{\cal D}}h(y)}.
\end{displaymath}
Alternatively, $h_{k-nn}(x)=\frac{1}{d^k(x)}$ with $d^k(x)$ being the 
dissimilarity to the
$k$th nearest neighbour would be another simple dissimilarity-based density
estimator, although this has no trivial upper bound ($h$, even before 
normalising by its within-cluster maximum, is bounded by $n$). One could also
standardise $h$ by the within-cluster maxima if clusters with generally lower
densities should have the same weight as high density clusters, 
but lower density values rely on fewer observations and are therefore less
reliable.

Three different aspects of density-based clustering are measured by three
different indexes:
\begin{enumerate}
\item The density should decrease within a cluster from the density mode to the
``outskirts'' of the cluster ($I_{densdec}$).
\item Cluster boundaries should run through density ``valleys'', i.e., 
high density points should not be close to many points from other clusters
($I_{densbound}$).
\item There should not be a big gap between high density regions within a
cluster ($I_{highdgap}$; gaps as measured by $I_{widestgap}$ may be fine in 
the low density outskirts of a cluster).
\end{enumerate}
The idea for $I_{densdec}$ is as follows. 
For every cluster, starting from the cluster mode,
i.e., the observation with the highest density, construct a growing sequence of 
observations that eventually covers the whole cluster by always adding the 
closest observation that is not yet included. Optimally, in this process,
the within-cluster density of newly included points should always decrease. 
Whenever actually the density goes up, a penalty of the squared
difference of the densities is incurred. 
The index $I_{densdec}$ aggregates these penalties. The following 
algorithm computes this, and it also constructs a set $T$ that collects 
information about high dissimilarities between high density observations and
is used for the definition of
$I_{highdgap}$ below:
\begin{description}
\item[Initialisation] $I_{d1}=0$, $T=\emptyset$. For $j=1,\ldots,K$: 
\item[Step 1] $S_j=\{x\}$, where $x=\argmax_{y \in C_j}h^*(y)$. 
\item[Step 2] Let $R_j=C_j\setminus S_j$. If $R_j=\emptyset$: $j=j+1$,
if $j\le K$ go to Step 1, if $j+K=1$ then go to Step 5. Otherwise:
\item[Step 3] Find $(x,y)=\argmin_{(z_1,z_2): z_1\in R_j, z_2\in S_j}d(z_1,z_2)$.
$S_j=S_j\cup\{x\}$, $T=T\cup \{\max_{z\in R_j}h^*(z)d(x,y)\}$. 
\item[Step 4] If $h^*(x)>h^*(y):\ I_{d1}=I_{d1}+(h^*(x)-h^*(y))^2$, back to Step 2.
\item[Step 5] $I_{densdec}({\cal C})=\sqrt{\frac{I_{d1}}{n}}.$   
\end{description}
$I_{densdec}$ collects the penalties 
from increases of the within-cluster densities 
during this process. 

The definition of $I_{densdec}$ 
does not take into account whether the neighbouring
observations that produce high density
values $h^*(x)$ for $x$ are in the same cluster as $x$. But this is important,
because it would otherwise be easy to achieve a good value of $I_{densdec}$ by 
cutting through high density areas and distributing a single high density 
area to several clusters.

A second index can be
defined that penalises a high contribution of points from different clusters
to the density values in a cluster (measured by $h_o$ below), 
because this means that the cluster border
cuts through a high density region.
\begin{displaymath}
\mbox{For }x_i,\ i=1,\ldots,n:\  h_o(x_i)=\sum_{k=1}^n \kappa(d(x_i,x_k))
1(\gamma(k)\neq\gamma(i)).
\end{displaymath}
Normalising:
\begin{displaymath}
h_o^*(x)=\frac{h_o(x)}{\max_{y\in {\cal D}} h(y)}.
\end{displaymath}
A penalty is incurred if for observations with a large density $h^*(x)$
there is a large contribution $h^*_o(x)$ to that density from other clusters:
\begin{displaymath}
  I_{densbound}({\cal C})=\frac{1}{n}\sum_{j=1}^K\sum_{x\in C_j} h^*(x)h^*_o(x).
\end{displaymath}
Both $I_{densdec}$ and $I_{densbound}$ are by definition $\ge 0$. Also, the maximum 
contribution of any observation to any of $I_{densdec}$ and $I_{densbound}$ is 
$\frac{1}{n}$, 
because the normalised $h^*$-values are $\le 1$. 
These are penalties, so
low values are good, and normalised versions are defined as
\begin{displaymath}
  I_{densdec}^*({\cal C})=1-I_{densdec}({\cal C}),\ 
  I_{densbound}^*({\cal C})=1-I_{densbound}({\cal C}).
\end{displaymath}
An issue with $I_{densdec}$ is that it is possible that there
is a large gap between two observations with high density, which does not
incur penalties if there are no low-density observations in between. This can be
picked up by a version of $I_{widestgap}$ based on the density-weighted gap 
information collected in $T$ above. This is suggested instead of
$I_{widestgap}$ if a density-based cluster concept is of interest:
\begin{displaymath}
  I_{highdgap}({\cal C})=\max T.  
\end{displaymath}
$I_{highdgap}({\cal C})\in[0,d_{max}]$ and low values are good, 
so it is normalised as 
$I^*_{highdgap}({\cal C})=1-\frac{I_{highdgap}({\cal C})}
{d_{max}}\in[0,1]$.

\subsection{Uniform within-cluster density}
Sometimes different clusters should not (only) be characterised by gaps between
them; overlapping regions in data space may be seen as different clusters if
they have different within-cluster density levels, which in some applications
could point to different data generating mechanisms behind the different 
clusters, which the researcher would like to discover. Such a cluster concept 
would require that densities within clusters are more or less uniform.

This can be characterised by the coefficient of variation CV of either the 
within-cluster density values or the dissimilarities to the $k$th nearest 
within-cluster neighbour $d^k_w(x)$ (say $k=4$). 
The latter is preferred here because 
as opposed to the density values, $d^k_w(x)$ is clean from the influence of 
observations from the other clusters. Define for $j=1,\ldots,k$, assuming 
$n_j>k$:
\begin{displaymath}
  m(C_j;k)=\frac{1}{n_j}\sum_{x\in C_j}d^k_w(x),\qquad  
 \CV(C_j)=\frac{\sqrt{\frac{1}{n_j-1}\sum_{x\in C_j}(d^k_w(x)-m(C_j;k))^2}}
{m(C_j;k)}.
\end{displaymath}
Using this,
\begin{displaymath}
  I_{cvdens}({\cal C})=\frac{\sum_{j=1}^K n_j \CV(C_j)1(n_j>k)}
{\sum_{j=1}^K  n_j1(n_j>k)}.
\end{displaymath}
Low values are good.
The maximum value of the coefficient of variation based on $n$ observations
is $\sqrt{n}$ (Katsnelson and Kotz\cite{KK57}), so a normalised version is 
$I_{cvdens}^*({\cal C})=1-\frac{I_{cvdens}({\cal C})}{\sqrt{n}}$.
\subsection{Entropy}
In some clustering applications, particularly where clustering is done for
``organisational'' reasons such as information compression, it is useful 
to have clusters that are roughly of the same size. This can be measured
by the entropy:
\begin{displaymath}
  I_{entropy}({\cal C})=-\sum_{j=1}^K \frac{n_j}{n}\log\left(\frac{n_j}{n}\right).
\end{displaymath}
Large values are good. The entropy is maximised for fixed $K$ by 
$e_{max}(K)=-\log\left(\frac{1}{K}\right)$, so it can be normalised
by $I_{entropy}^*({\cal C})=\frac{I_{entropy}({\cal C})}{e_{max}(K)}$. 
\subsection{Parsimony}
In case that there is a preference for a lower number of clusters, one could
simply define 
\begin{displaymath}
  I_{parsimony}^*=1-\frac{K}{K_{max}},
\end{displaymath}
(already normalised) with $K_{max}$ the maximum number of clusters of 
interest. If in a given 
application there is a known nonlinear loss connected to the number of 
clusters, this can obviously be used instead, and the principle can be
applied also to other free parameters of a clustering method, if desired. 
\subsection{Similarity to homogeneous distributional shapes}
Sometimes the meaning of ``homogeneity'' for a cluster is defined by 
a homogeneous probability model, e.g., Gaussian mixture model-based clustering
models all clusters by Gaussian distributions with different parameters,
requiring Euclidean data. 
Historically, due to the Central Limit Theorem and Quetelet's 
``elementary error hypothesis'', measurement errors were widely believed to
be normally/Gaussian distributed (see Stigler\cite{Sti86}). 
Under such a hypothesis it makes sense 
in some situations to regard Gaussian distributed observations as homogeneous,
and as pointing to the same underlying mechanism; this could also motivate to
cluster observations together that look like being generated from the same 
(approximate) Gaussian distribution. Indexes that measure cluster-wise 
Gaussianity can be defined, see, e.g., Lago-Fernandez and Corbacho\cite{LFC10}. 
One possible principle is to compare a 
one-dimensional function of the observations within a cluster to its theoretical
distribution under the data distribution of interest; e.g., Coretto and 
Hennig\cite{CH16} compare the Mahalanobis distances of observations to their 
cluster centre with their theoretical $\chi^2$-distribution using the
Kolmogorow-distance. This is also possible for other distributions of interest.
\subsection{Stability}
Clusterings are often interpreted as meaningful in the sense that they can be
generalised as substantive patterns. This at least implicitly requires that
they are stable. Stability in cluster analysis can be explored using resampling
techniques such as bootstrap and splitting the dataset, and clustering from 
different resampled datasets can be compared. This requires to run the 
clustering method again on the resampled datasets and I will not treat this
here in detail, but useful indexes have been defined using this principle,
see, e.g., Tibshirani and Walther\cite{TW05} and Fang and Wang\cite{FW12}.
\subsection{Further Aspects}
Hennig\cite{Hen16} lists further potentially desirable characteristics of a 
clustering, for which further indexes could be defined:
\begin{itemize}
\item Areas in data space corresponding to clusters should have certain
characteristics such as being linear or convex.
\item It should be possible to characterise clusters using a small number
of variables.
\item Clusters should correspond well to an externally given partition or
values of an external variable (this could for example imply that clusters of
regions should be spatially connected).
\item Variables should be approximately independent within clusters.
\end{itemize}

\section{Aggregation of indexes} \label{sagg}
The required cluster concept and therefore the way the validation indexes can be used depends on the specific clustering application. The users need to specify 
what characteristics of the clustering are desired in the application. The 
corresponding indexes can then be aggregated to form a single criterion that 
can be used to compare different clustering methods, different numbers of 
clusters and other possible parameter choices of the clustering.

The most straightforward aggregation is to compute a weighted mean of $s$ 
selected indexes $I_1,\ldots,I_s$ with weights $w_1,\ldots,w_s> 0$ 
expressing the relative importance of the different methods:
\begin{equation}\label{eagg}
  A({\cal C})=\sum_{k=1}^s w_kI_k.
\end{equation}
Assuming that large values are desirable for all of $I_1,\ldots,I_s$, the best
clustering for the application in question can be found by maximising $A$. 
This can be done by comparing different clusterings from conventional clustering
methods, but in principle it would also be an option to try to optimise $A$
directly.

The weights can only be chosen to directly reflect the relative importance of 
the various aspects of a clustering if the values (or, more precisely, their
variations) of the indexes $I_1,\ldots,I_s$ are comparable, and give the indexes
equal influence on $A$ if all weights are equal. In Section 
\ref{saspects} I proposed tentative normalisations of all indexes, which 
give all indexes the same value range $[0,1]$. Unfortunately this is not 
good enough to ensure comparability; on many datasets some of these indexes
will cover almost the whole value range whereas other indexes may be larger 
than 0.9 for all clusterings that any clustering method would come up with.  
Therefore,
Section \ref{sstupid} will introduce a new computational method to standardise
the variation of the different criteria.

Another issue is that some indexes by their very nature favour large numbers
of clusters $K$ (obviously large within-cluster dissimilarities can be more 
easily avoided for large $K$), whereas others favour small values of $K$
(separation is more difficult to achieve with many small clusters). The method
introduced in Section \ref{sstupid} will allow to assess the extent to which
the indexes deliver systematically larger or smaller values for larger $K$.
Note that this can also be an issue for univariate ``global'' 
validation indexes from the literature, see Hennig and Lin\cite{HL15}.
 
If the indexes should be used 
to find an optimal value of $K$, the indexes in $A$ 
should be chosen in such a way that 
indexes that systematically favour larger $K$ and indexes that systematically
favour smaller $K$ are balanced. 

The user needs to take into account that 
the proposed indexes are not independent.
For example, good representation of objects by centroids will normally be 
correlated with having generally small within-cluster dissimilarities. Including
both indexes will assign extra weight to the information that the two indexes
have in common (which may sometimes but not always be desired). 

There are alternative ways to aggregate the information from the different
indexes. For example, one could use some indexes as side conditions rather than
involving them in the definition of $A$. For example, rather than giving
entropy a weight for aggregation as part of $A$, one may specify a certain 
minimum entropy value below which clusterings are not accepted, but not
use the entropy value to distinguish between clusterings that fulfil the
minimum entropy requirement. Multiplicative aggregation is another option.


\section{Random clusterings for calibrating indexes} \label{sstupid}
As explained above, the normalisation in Section \ref{saspects} does not provide
a proper calibration of the validation indexes. Here is an idea for doing this 
in a more appropriate way. The idea is that random clusterings are generated
on ${\cal D}$ and index values are computed, in order to explore what range of
index values can be expected on ${\cal D}$, so that the clusterings of interest
can be compared to these. So in this Section, as opposed to conventional 
probability modelling, the dataset is considered as fixed 
but a distribution of index values is generated from various random partitions.

Completely random clusterings (i.e., assigning every observation independently 
to a cluster) are not suitable for this, because it can be expected that indexes
formalising desirable characteristics of a clustering will normally give much 
worse values for them than for clusters that were generated by a clustering
method. Therefore I propose two methods for random clusterings that are meant 
to generate clusterings that make some sense, at least by being connected in 
data space. The methods are called ``stupid K-centroids'' and ``stupid nearest 
neighbours''; ``stupid'' because they are versions of popular clustering 
methods (centroid-based clustering like K-means or PAM, and 
Single Linkage/Nearest Neighbour) that replace optimisation by random decisions
and are meant to be computable very quickly. Centroid-based clustering normally
produces somewhat compact clusters, whereas Single Linkage is notorious for 
prioritising cluster separation totally over within-cluster homogeneity, and 
therefore one should expect these two approaches to explore in a certain 
sense opposite ways of clustering the data.

\subsection{Stupid K-centroids clustering}
Stupid K-centroids works as follows. For fixed number of cluster $K$ draw 
a set of $K$ cluster centroids $Q=\{q_1,\ldots,q_K\}$ from ${\cal D}$ 
so that every subset
of size $K$ has the same probability of being drawn. 
${\cal C}_{K-stupidcent}(Q)=\{C_1,\ldots,C_k\}$ is defined by assigning every
observation to the closest centroid: 
\begin{displaymath}
  \gamma(i)=\argmin_{j\in\{1,\ldots,K\}} d(x_i,q_j),\ i=1,\ldots,n.
\end{displaymath}
\subsection{Stupid nearest neighbours clustering}
Again, for fixed number of cluster $K$ draw 
a set of $K$ cluster initialisation points 
$Q=\{q_1,\ldots,q_K\}$ from ${\cal D}$ 
so that every subset
of size $K$ has the same probability of being drawn.
${\cal C}_{K-stupidnn}(Q)=\{C_1,\ldots,C_k\}$ is defined by successively
adding the not yet assigned observation closest to any cluster to that cluster
until all observations are clustered:
\begin{description}
\item[Initialisation] Let $Q^*=Q$. Let 
\begin{displaymath}
{\cal C}^*(Q)={\cal C}^*(Q^*)=\{C_1^*,\ldots,C_L^*\}=\left\{\{q_1\},\ldots,\{q_K\}\right\}.
\end{displaymath}
\item[Step 1] Let $R^*={\cal D}\setminus Q^*$. If $R^*\neq\emptyset$, 
find $(x,y)=\argmin_{(z,q): z\in R^*, q\in Q^*}d(z,q)$, otherwise stop. 
\item[Step 2] Let $Q^*=Q^*\cup \{x\}$.
For the $C^*\in {\cal C}^*(Q^*)$ with $y\in C^*$, let 
$C^*=C^*\cup \{x\}$, updating ${\cal C}^*(Q^*)$ accordingly. Go back to Step 1.
\end{description}
At the end, ${\cal C}_{K-stupidnn}(Q)={\cal C}^*(Q^*)$. 
\subsection{Calibration}
The random clusterings can be used in various ways to calibrate the indexes. 
For any 
value $K$ of interest,
$2B$ clusterings ${\cal C}_{K-collection}=({\cal C}_{K:1},\ldots,{\cal C}_{K:2B})=$
\begin{displaymath}
\left({\cal C}_{K-stupidcent}(Q_1),\ldots,{\cal C}_{K-stupidcent}(Q_B),
{\cal C}_{K-stupidnn}(Q_1),\ldots,{\cal C}_{K-stupidnn}(Q_B)\right)
\end{displaymath}
on ${\cal D}$ are generated, say $B=100$. 

As mentioned before, indexes may systematically change over $K$ and therefore 
may show a preference for either large or small $K$. In order to account for 
this, it is possible to calibrate the indexes using stupid clusterings for the
same $K$, i.e., for a clustering ${\cal C}$ with $|{\cal C}|=K$. Consider
an index $I^*$ of interest (the normalised version is used
here because this means that large values are good for all indexes). Then,
\begin{equation}\label{ecali}
  I^{cK}({\cal C})=\frac{I^*({\cal C})-m^*({\cal C}_{K-collection})}
{\sqrt{\frac{1}{2B-1}\sum_{j=1}^{2B}\left(I^*({\cal C}_{K:j})-m^*({\cal C}_{K-collection})\right)^2}},
\end{equation}
where $m^*({\cal C}_{K-collection})=\frac{1}{2B}\sum_{j=1}^{2B}I^*({\cal C}_{K:j})$.
A desired set of calibrated indexes can then be used for aggregation in 
(\ref{eagg}). 

An important alternative to (\ref{ecali}) is calibration by using random 
clusterings for all values of $K$ together. Let 
${\cal K}=\{2,\ldots,K_{max}\}$ be the numbers of clusters of interest (most
indexes will not work for $K=1$), 
${\cal C}_{collection}=\{C_{K:j}:\ K\in{\cal K},\ j=1,\ldots,2B\}$, 
$m^*({\cal C}_{collection})=\frac{1}{2B(K_{max}-1)}\sum_{K=2}^{K_{max}}
\sum_{j=1}^{2B}I^*({\cal C}_{K:j})$. With this,
\begin{equation}\label{eallk}
  I^{c}({\cal C})=\frac{I^*({\cal C})-m^*({\cal C}_{collection})}
{\sqrt{\frac{1}{2B(K_{max}-1)-1}\sum_{K=2}^{K_{max}}
\sum_{j=1}^{2B}\left(I^*({\cal C}_{K:j})-m^*({\cal C}_{collection})\right)^2}}.
\end{equation}
$I^c$ does not correct for potential systematic tendencies of the indexes
over ${\cal K}$, but this is not a problem if the user is happy to use the
uncalibrated indexes directly for 
comparing different values of $K$; a potential bias toward large or small values
of $K$ in this case needs to be addressed by choosing the indexes to be 
aggregated in (\ref{eagg}) in a balanced way. This can be checked by computing
the aggregated index $A$ also for the random clusterings and check how these
change over the different values of $K$.
 
Another  
alternative is to calibrate indexes by using their rank value in the
set of clusterings (random clusterings and clusterings to compare) rather than
a mean/standard deviation-based standardisation. This is probably more robust 
but comes with some loss of information.

\section{Examples}\label{sexamples}
{\bf Withindis needs recomputing here because of reweighting!}
\subsection{Artificial dataset}
The first example is the artificial dataset shown in Figure \ref{fxyunif}. 
Four clusterings are compared (actually many more clusterings with $K$ between
2 and 5 were compared on these data, but the selected clusterings illustrate the
most interesting issues).

The clusterings were computed by K-means with $K=2$ and $K=3$, Single Linkage
cut at $K=3$ and PAM with $K=5$. The K-means clustering with $K=3$ and the 
Single Linkage clustering are shown in Figure \ref{fxyunif}. The K-means
clustering with $K=2$ puts the uniformly distributed widespread point cloud
on top together in a single cluster, and the two smaller populations are the 
second cluster. This is the most intuitive clustering for these data for $K=2$ 
and also delivered by most other clustering methods. PAM does not separate 
the two smaller (actually Gaussian) populations for $K=2$, but it does so for 
$K=5$, along with splitting the uniform point cloud into three parts. 

\begin{table}
  \begin{center}
    \begin{tabular}{|l|rrrr|}
      \hline
       & kmeans-2 & kmeans-3 & Single Linkage-3 & PAM-5\\
      \hline
       $I_{withindis}^*$ &  0.654   & 0.799 & 0.643 & 0.836\\
       $I_{0.1-sep}^*$ &  0.400   & 0.164 & 0.330 & 0.080\\
       $I_{centroid}^*$ &  0.766   & 0.850 & 0.790 & 0.896\\
       $I_{Pearson\Gamma}^*$ &  0.830   & 0.900 & 0.781 & 0.837\\
       $I_{widestgap}^*$ &  0.873   & 0.873 & 0.901 & 0.901\\
       $I_{densdec}^*$ &  0.977   & 0.981 & 0.981 & 0.985\\
       $I_{densbound}^*$ &  1.000   & 0.999 & 1.000 & 0.997\\
       $I_{highdgap}^*$ &  0.879   & 0.879 & 0.960 & 0.964\\
       $I_{cvdens}^*$ &  0.961   & 0.960 & 0.961 & 0.959\\
       $I_{entropy}^*$ &  0.863   & 0.993 & 0.725 & 0.967\\
      \hline
    \end{tabular}
  \end{center}
  \caption{Normalised index values for four clusterings on artificial data.}
  \label{txyunifraw}
\end{table}

Table \ref{txyunifraw} shows the normalised index values for these clusterings.
Particularly comparing 3-means and Single Linkage, the different virtues of
these clusterings are clear to see. 3-means is particularly 
better for the homogeneity-driven $I_{withindis}^*$ and $I_{centroid}^*$, whereas 
Single Linkage wins 
regarding the separation-oriented $I_{0.1-sep}^*$ and $I_{widestgap}^*$, with 
3-means ignoring the gap between the two Gaussian populations. 
$I_{Pearson\Gamma}^*$ tends toward
3-means, too, which was perhaps less obvious, because it does not like too big 
distances within clusters. It is also preferred by $I_{entropy}^*$ because
of joining two subpopulations that are rather small. 
The values for the  indexes, $I_{densdec}^*$, $I_{densbound}^*$, $I_{highdgap}^*$, 
and $I_{cvdens}^*$
illustrate that that the naive normalisation is not quite suitable for making 
the value ranges of the indexes comparable. For the 
density-based indexes, many involved terms are far away from the maximum used
for normalisation, so the index values can be close to 0 (close to 1 after
normalisation). This is amended by calibration.

Considering the clusterings with $K=2$ and $K=5$, it can be seen that with $K=5$
it is easier to achieve within-cluster homogeneity ($I_{withindis}^*$, 
$I_{centroid}^*$), whereas with $K=2$ it is easier to achieve separation 
($I_{0.1-sep}^*$). 

\begin{table}
  \begin{center}
    \begin{tabular}{|l|rrrr|}
      \hline
       & kmeans-2 & kmeans-3 & Single Linkage-3 & PAM-5\\
      \hline
       $I_{withindis}^{cK}$ &  0.906   & 1.837 & -0.482 & 0.915\\
       $I_{0.1-sep}^{cK}$ &  1.567   & 0.646 & 3.170 & -0.514\\
       $I_{centroid}^{cK}$ &  1.167   & 1.599 & 0.248 & 1.199\\
       $I_{Pearson\Gamma}^{cK}$ &  1.083   & 1.506 & 0.099 & 0.470\\
       $I_{widestgap}^{cK}$ &  1.573   & 1.156 & 1.364 & 0.718\\
       $I_{densdec}^{cK}$ &  1.080   & 1.191 & 1.005 & 1.103\\
       $I_{densbound}^{cK}$ &  0.452   & 0.449 & 0.519 & 0.647\\
       $I_{highdgap}^{cK}$ &  1.317   & 0.428 & 2.043 & 1.496\\
       $I_{cvdens}^{cK}$ &  1.153   & 0.836 & 0.891 & 0.286\\
       $I_{entropy}^{cK}$ &  0.246   & 1.071 & -0.620 & 0.986\\
      \hline
    \end{tabular}
  \end{center}
  \caption{Calibrated index values (using random clusterings with same $K$) for four clusterings on artificial data.}
  \label{txyunifck}
\end{table}

Table \ref{txyunifck} shows the index values $I^{cK}$ calibrated against 
random clustering with the same $K$. This is meant to account for the fact that
some indexes differ systematically over different values of $K$. Indeed, using
this calibration, PAM with $K=5$ is no longer best for $I_{centroid}^{cK}$ and
$I_{withindis}^{cK}$, and
2-means is no longer best for $I_{0.1-sep}^{cK}$. It can now be seen that 
3-means is better than Single Linkage for $I_{densdec}^{cK}$. This is because 
density values show much more variation in the widely spread uniform 
subpopulation than in the two small Gaussian ones, so splitting up the uniform
subpopulation is better for creating densities decreasing from the modes, 
despite the gap between the two Gaussian subpopulations. On the other hand, 
3-means has to cut through the uniform population, which gives Single Linkage,
which only cuts through clear gaps, an advantage regarding $I_{densbound}^{cK}$,
and particularly 3-means incurs a large distance between the
two Gaussian high density subsets within one of its clusters, which makes
Single Linkage much better regarding $I_{highdgap}^{cK}$. 
Ultimately, the user needs to decide here whether small within-cluster
dissimilarities and short dissimilarities to centroids
are more important than separation and the absence of within-cluster gaps.
The $K=5$-solution does not look very attractive regarding most criteria
(although calibration with the same $K$ makes it look good regarding
$I_{densbound}^{cK}$); the $K=2$-solution only 
looks good regarding two criteria that
may not be seen as the most important ones here.

\begin{table}
  \begin{center}
    \begin{tabular}{|l|rrrr|}
      \hline
       & kmeans-2 & kmeans-3 & Single Linkage-3 & PAM-5\\
      \hline
       $I_{withindis}^{c}$ & -0.483   & 1.256 & -0.607 & 1.694\\
       $I_{0.1-sep}^{c}$ &  2.944   & 0.401 & 2.189 & -0.512\\
       $I_{centroid}^{c}$ & -0.449   & 0.944 & -0.059 & 1.712\\
       $I_{Pearson\Gamma}^{c}$ &  0.658   & 1.515 & 0.058 & 0.743\\
       $I_{widestgap}^{c}$ &  0.939   & 0.939 & 1.145 & 1.145\\
       $I_{densdec}^{c}$ & -0.279   & 0.832 & 0.697 & 1.892\\
       $I_{densbound}^{c}$ &  0.614   & 0.551 & 0.609 & 0.417\\
       $I_{highdgap}^{c}$ &  0.464   & 0.464 & 1.954 & 2.025\\
       $I_{cvdens}^{c}$ &  0.761   & 0.692 & 0.748 & 0.615\\
       $I_{entropy}^{c}$ &  0.208   & 1.079 & -0.720 & 0.904\\
      \hline
    \end{tabular}
  \end{center}
  \caption{Calibrated index values (using all random clusterings) for four clusterings on artificial data.}
  \label{txyunifc}
\end{table}

Table \ref{txyunifc} shows the index values $I^{cK}$ calibrated against 
all random clusterings. Not much changes regarding the comparison of 3-means
and Single Linkage, whereas a user who is interested in small within-cluster
dissimilarities and centroid representation in absolute terms is now drawn
toward PAM with $K=5$ or even much larger $K$, indicating that these indexes
should not be used without some kind of counterbalance, either from 
separation-based criteria ($I_{0.1-sep}^{c}$ and $I_{densbound}^{c}$) 
or taking into account parsimony. A high density gap within a cluster
is most easily avoided with large $K$, too, whereas $K=2$ achieves the best 
separation, unsurprisingly.
 
As this is an artificial dataset and there is no subject-matter information that
could be used to prefer certain indexes, I do not present specific 
aggregation weights here.

\subsection{Tetragonula bees data}
Franck \textit{et al.}\cite{FCGRO04} published a data set giving genetic 
information about
236 Australasian tetragonula bees, in which it is of interest to determine the
number of species. The data set is incorporated in the package ``fpc'' of the
software system R (\verb|www.r-project.org|) and is available on the IFCS 
Cluster Benchmark Data Repository \verb|http://ifcs.boku.ac.at/repository|. 
Bowcock \textit{et al.}\cite{BRTMKC94} defined the
``shared allele dissimilarity'' formalising genetic dissimilarity 
appropriately for
species delimitation, which is used for the present data set. It yields values 
in $[0, 1]$. See also Hausdorf and Hennig\cite{HH10} and Hennig\cite{Hen13}
for earlier analyses of this dataset including a discussion of the number of
clusters problem. Franck \textit{et al.}\cite{FCGRO04} provide 9 
``true'' species for these data, although this manual 
classification (using morphological information besides genetics) 
comes with its own problems and may not be 100\% reliable.
 
In order to select indexes and to find weights, some knowledge about species
delimitation is required, which was provided by Bernhard Hausdorf, Museum of 
Zoology, University of Hamburg. The biological 
species concept requires that there is 
no (or almost no) genetic exchange between different species, so that separation
is a key feature for clusters that are to be interpreted as species. For the
same reason, large within-cluster gaps can hardly be tolerated (regardless of
the density values associated to them); in such a case
one would consider the subpopulations on two sides of a gap separate species, 
unless a case can be made that potentially existing connecting individuals
could not be sampled. Gaps may also occur in regionally separated subspecies,
but this cannot be detected from the data without regional information.
On the other hand, species should be reasonably homogeneous; it would be 
against biological intuition to have strongly different genetic patterns within
the same species. This points to the indexes $I_{withindis}$, $I_{0.1-sep}$, and
$I_{widestgap}$. On the other hand, the shape of the within-cluster density is
not a concern here, and neither are representation of clusters by centroids,
entropy, and constant within-cluster variation. The index  $I_{Pearson\Gamma}$ 
is added to the set of relevant indexes, because one can interpret the species
concept as a representation of genetic exchange as formalised by the shared
allele dissimilarity, and $I_{Pearson\Gamma}$ measures the quality of this 
representation. All these four indexes are used in (\ref{eagg}) with weight 1
(one could be interested in stability as well, which is not taken into account
here). 

\begin{table}
  \begin{center}
    \begin{tabular}{|l|rrrrrrrr|}
      \hline
       & AL-5  & AL-9 & AL-10 & AL-12 & PAM-5 & PAM-9 & PAM-10 & PAM-12 \\
      \hline
 $I_{withindis}^{cK}$ & 0.68 & -0.04 & 1.70 & 1.60 & 1.83 & 2.45 & 2.03 & 1.80\\
 $I_{0.1-sep}^{cK}$ &  1.79  & 2.35 & 2.00 & 2.42 & 0.43 & 1.59 & 2.12 & 0.94\\
 $I_{Pearson\Gamma}^{cK}$ & 1.86 & 2.05 & 1.92 & 2.28 & 1.43 & 1.84 & 1.75 & 0.61\\
 $I_{widestgap}^{cK}$ & 0.45 & 4.73 & 4.90 & 4.86 & -1.03 & 0.41 & 0.42 & -0.09\\
 $A({\cal C})$ & 4.78  & 9.09 & 10.51 & 11.13 & 2.66 & 6.30 & 6.32 & 3.30\\
 ARI           &  0.53 & 0.60 & 0.95 & 0.94 & 0.68 & 0.84 & 0.85 & 0.64\\
      \hline
    \end{tabular}
  \end{center}
  \caption{Calibrated index values (using random clusterings with same $K$) for eight clusterings on Tetragonula bees data with aggregated index and adjusted Rand index.}
  \label{ttetrack}
\end{table}

Again I present a subset of the clusterings that were actually compared 
for illustrating the use of the approach presented in this paper. Typically
clusterings below $K=9$ were substantially different from the ones with 
$K\ge 9$; clusterings with $K=10$ and $K=11$ from the same method were often 
rather similar to each other, and I present clusterings from Average Linkage
and PAM with $K=5, 9, 10$, and 12. Table \ref{ttetrack} shows the 
four relevant index values $I^{cK}$ calibrated against random clustering with 
the same $K$ along with the aggregated index $A({\cal C})$. Furthermore, 
the adjusted Rand index (ARI; Hubert and Arabie\cite{HA85}) comparing the
clusterings from the method with the ``true'' species is given (this takes 
values between -1 and 1 with 0 expected for random clusterings and 1 for
perfect agreement). Note that despite $K=9$ being the number of ``true'' 
species, clusterings with $K=10$ and $K=12$ yield higher ARI-values than those
with $K=9$, so these clusterings are preferable (it does not help much to 
estimate the number of species correctly if the species are badly composed).
Some ``true'' species in the original dataset are widely regionally dispersed 
with hardly any similarity between subspecies.

The aggregated index
$A({\cal C})$ is fairly well related to the ARI (over all 55 clusterings 
that were compared the correlation between $A({\cal C})$ and ARI 
is about 0.85). The two clusterings that are
closest to the ``true'' one also have the highest values of $A({\cal C})$.
The within-cluster gap criterion plays a key role here, preferring Average
Linkage with 9-12 clusters clearly over the other clusterings.
$A({\cal C})$ assigns its highest value to AL-12, whereas the ARI for AL-10 is
very slightly higher. PAM delivers better clusterings regarding small 
within-cluster dissimilarities, but this advantage is dwarfed by the advantage 
of Average Linkage regarding separation and within-cluster gaps.

\begin{table}
  \begin{center}
    \begin{tabular}{|l|rrrrrrrr|}
      \hline
       & AL-5  & AL-9 & AL-10 & AL-12 & PAM-5 & PAM-9 & PAM-10 & PAM-12 \\
      \hline
 $I_{withindis}^{c}$ & 0.10 & 0.59 & 1.95 & 2.00 & 0.83 & 2.13 & 2.17 & 2.16\\
 $I_{0.1-sep}^{c}$ &  1.98  & 1.54 & 1.05 & 1.02 & 0.53 & 1.01 & 1.13 & 0.21\\
 $I_{Pearson\Gamma}^{c}$ & 1.79 & 1.87 & 1.86 & 1.87 & 1.38 & 1.71 & 1.73 & 0.72\\
 $I_{widestgap}^{c}$ & 0.39 & 5.08 & 5.08 & 5.08 & -1.12 & 0.39 & 0.39 & -0.08\\
 $A({\cal C})$ & 4.26  & 9.08 & 9.93 & 9.97 & 1.62 & 5.24 & 5.41 & 3.01\\
 ARI           &  0.53 & 0.60 & 0.95 & 0.94 & 0.68 & 0.84 & 0.85 & 0.64\\
      \hline
    \end{tabular}
  \end{center}
  \caption{Calibrated index values (using all random clusterings) for eight clusterings on Tetragonula bees data with aggregated index and adjusted Rand index.}
  \label{ttetrac}
\end{table}

Table \ref{ttetrac} shows the corresponding results with calibration using all
random clusterings. This does not result in a different ranking of the 
clusterings, so this dataset does not give a clear hint which of the two 
calibration methods is more suitable, or, in other words, the results do not
depend on which one is chosen.

\section{Conclusion}\label{sconc}
The multivariate array of cluster validation indexes presented here provides
the user with a detailed characterisation of various relevant aspects of a 
clustering. The user can aggregate the indexes in a suitable way to find a 
useful clustering for the clustering aim at hand. 

The indexes can also be used to provide a more detailed comparison of different
clustering methods in benchmark studies, and a better understanding of their
characteristics.

The methodology is currently partly implemented in the ``fpc''-package of
the statistical software system R and will soon be fully implemented there.

Most indexes require $K\ge 2$ and the approach can therefore not directly be
used for deciding whether the dataset is homogeneous as a whole ($K=1$). The
individual indexes as well as the aggregated index could be used in a 
parametric bootstrap scheme as proposed by Hennig and Lin\cite{HL15} to test
the homogeneity null hypothesis against a clustering alternative.

Research is still required in order to compare the different calibration 
methods and some alternative versions of indexes. A theoretical characterisation
of the indexes is of interest as well as a study exploring the strength of the
information overlap between some of the indexes, looking at, e.g., correlations
over various clusterings and datasets. Random clustering calibration may also be
used together with traditional univariate validation indexes. 
Further methods for random clustering could be developed and it could
be explored what
collection of random clusterings is most suitable for calibration (some work in
this direction is currently done by my PhD student Serhat Akhanli).

\section*{Acknowledgement}
This work was supported by EPSRC Grant EP/K033972/1.

\end{document}